# Understanding University Students' Use of Generative AI: The Roles of Demographics and Personality Traits

Newnew Deng[1*][0009-0006-8486-0391], Edward Jiusi Liu[1*][0009-0009-4649-2032], Xiaoming Zhai[2**]

[1] West Lafayette Junior/Senior High School, West Lafayette, IN 47906, USA
[2] University of Georgia, Athens, GA 30666, USA
*Co-first Author, Alphabetically Listed
** Corresponding Author

xiaoming.zhai@uga.edu

**Abstract.** The use of generative AI (GAI) among university students is rapidly increasing, yet empirical research on students' GAI use and the factors influencing it remains limited. To address this gap, we surveyed 363 undergraduate and graduate students in the United States, examining their GAI usage and how it relates to demographic variables and personality traits based on the Big Five model (i.e., extraversion, agreeableness, conscientiousness, and emotional stability, and intellect/imagination). Our findings reveal: (a) Students in higher academic years are more inclined to use GAI and prefer it over traditional resources. (b) Non-native English speakers use and adopt GAI more readily than native speakers. (c) Compared to White, Asian students report higher GAI usage, perceive greater academic benefits, and express a stronger preference for it. Similarly, Black students report a more positive impact of GAI on their academic performance. Personality traits also play a significant role in shaping perceptions and usage of GAI. After controlling demographic factors, we found that personality still significantly predicts GAI use and attitudes: (a) Students with higher conscientiousness use GAI less. (b) Students who are higher in agreeableness perceive a less positive impact of GAI on academic performance and express more ethical concerns about using it for academic work. (c) Students with higher emotional stability report a more positive impact of GAI on learning and fewer concerns about its academic use. (d) Students with higher extraversion show a stronger preference for GAI over traditional resources. (e) Students with higher intellect/imagination tend to prefer traditional resources. These insights highlight the need for universities to provide personalized guidance to ensure students use GAI effectively, ethically, and equitably in their academic pursuits.

**Keywords:** Generative AI (GAI), University Students, Personality Trait, Academic

## 1      Introduction

University students utilize generative AI (GAI) for various academic purposes, such as learning, assignments, essays, programming, and examinations, with differing levels of engagement and embracement [18,20]. However, empirical evidence is needed to examine the uses and perceptions of GAI in academic work, as well as the influencing factors. In this study, we surveyed university students to understand not only how they integrate GAI into their academic work but also how their use of GAI is associated with student demographics and personality traits. We aim to gain a comprehensive understanding of the factors that drive or hinder GAI adoption in academia so that policymakers, educators, and technology developers can make informed decisions when developing guidelines, interventions, and tools to better support students' learning.

## 2      Use of GAI and Influencing Factors

University students are increasingly reported to use GAI for various academic work. Research has examined students' use of ChatGPT and its effectiveness in various learning tasks. Some studies have shown that GAI improves students' learning and achievements, such as programming [7], math learning [4], writing [11], English learning [8], and lab tasks [2]. For example, Guhan et al. surveyed Engineering English as a Second Language students regarding their experiences and perceptions of employing ChatGPT in the English language classroom. Overall, the students reported that ChatGPT has greatly enhanced learners' vocabulary, listening, speaking, writing, and reading skills [7]. On the other hand, concerns have been raised about the use of GAI in academic work (e.g., the accuracy of the content), as over-reliance might have a negative impact on students' critical thinking and problem-solving abilities, posing a threat to academic ethics [9].

### 2.1      Student Demographics and GAI Use

Researchers have explored how GAI use is associated with various factors. A five-country study on the factors influencing university students' attitudes and use of ChatGPT found that students' use of ChatGPT is positively associated with their perceived ease of use, perceived usefulness, positive attitude towards technology, social influence, behavioral and cognitive elements, low anxiety, and minimal perceived risks [1]. They suggested that policies for ChatGPT adoption in higher education should be tailored to individual contexts, considering the variations in student attitudes.

Zhang et al. (2024) investigated the association between socioeconomic status (SES) and digital and AI literacy with the types of Chat GPT use among college students. [21]. They measured students' SES using seven questions that inquired about first-generation university student status, percentage of student loans, parental education, estimated family income, perceived family social class, and socioeconomic status. They found that university students from higher socioeconomic status (SES) families tend to use ChatGPT more frequently. They suggested that future studies should investigate the gender and racial/ethnic divides in ChatGPT use.

Students' academic standing may also influence the use of GAI, with senior students engaging with it more frequently. Surveys found that approximately 46% of high school students use AI tools, while more than half of university students use AI tools, surpassing the proportion of faculty members in higher education [14].

## 2.2  The Influence of Personality Traits on GAI Use

In addition to demographic information, research has examined the impact of personality on GAI use, as personality is considered one of the most fundamental psychological characteristics that influence the cognitive, affective, and behavioral actions of individuals [16]. The Big Five Personality traits are widely recognized in psychology as a framework for describing human personality and understanding human behavior, encompassing five broad domains (e.g., [3, 15, 17]): (a) Extraversion — Being outgoing, energetic, and sociable. (b) Agreeableness — Being compassionate, cooperative, helpful, empathetic, and trusting. (c) Conscientiousness — being organized, responsible, disciplined, and goal-oriented., (d) Emotional Stability—Being calm, secure, and free from persistent negative feelings, distress, and mood swings. And (e) Intellect/Imagination — being imaginative, creative, curious, and open to unconventional ideas and values.

Faruk et al. (2023) employed the Big Five Personality traits to study the relationship between students' use of AI and their personalities [5]. They found that individuals who were open to learning had a high correlation with the use of ChatGPT. In contrast, traits such as neuroticism and agreeableness have a negative influence on students' perception of GAI usefulness. In contrast, Filippi (2024) found that agreeableness has a positive influence on students' perception of GAI, which in turn impacts students' acceptance of AI [6].

Researchers also examined the combined influence of demographic information and personality on GAI. For example, Kaya et al. (2024) analyzed the relationships between demographic information, personality, and attitudes toward AI among adults recruited online. After controlling for demographic information (age, gender, educational level), they found that higher computer use and greater knowledge of AI are significant positive predictors of positive general attitudes toward AI, as AI learning anxiety is also a significant negative predictor of positive attitudes. Agreeableness significantly predicted negative attitudes toward AI [13].

We further the research on GAI use and influencing factors in the following ways: First, we further explore the conflicting influence of personality on GAI use, given that personality has shown inconsistent effects on students' GAI use in the existing literature, such as the influence of agreeableness. Second, we recruited participants from a diverse range of university students, including GAI users and non-users, as well as undergraduate and graduate students. In contrast, Faruk et al. (2023) only recruited

ChatGPT users, excluding those who do not use the platform [5]. Kaya recruited adult participants online. Third, we examine how GAI use, and perception are associated with a range of demographic characteristics specific to college students, such as grade levels, majors, and language status — an under-investigated area —and the Big Five personality traits. This study aims to answer the following research questions:

1. How do university students use generative AI?
2. How is their demographic information—specifically gender, ethnicity, language status, and grade level—associated with GAI use?
3. How are students' personality traits related to their use and perceptions of GAI after controlling for significant demographic factors?

By answering these questions, the study provides valuable insights into the role of individual differences in GAI use and perception, enabling educators to better guide students in using GAI effectively and appropriately in higher education.

## 3   Methods

### 3.1   Participants

A total of 363 students from a public university in the USA participated in the study across three days in a university computer lab. After each student had signed the IRB consent form, they completed three surveys via Qualtrics in the lab and received $20 each as compensation for their time. The participants represented a diverse range of gender, ethnicity, academic major, grade level, and native English speaker status (Appendix A[1]).

### 3.2   Instruments

We developed five items to examine students' demographic information (Appendix A[1]) and 15 items to measure students' GAI use and perceptions (Appendix B[1]). To measure Big Five Personality, we adopted the 50-item International Personality Item Pool (IPIP) [10, 12], with 10 items measuring each trait. This instrument has been considered the most reliable for measuring the Big Five Personality traits, and its validity has also been examined and supported [19]. After reversing the scores for any reversed statements, we calculated the alpha coefficient for the items measuring each trait based on our data. The results show that all the alpha coefficients are higher than .70 (Table 1). We further created five composite scores for the Big Five traits by averaging all the items measuring each construct and used the composite scores for the analysis.

---

[1]   https://github.com/AI4STEM-Education-Center/UnderstandingUniversity-Students-Use-of-Generative-AI

Table 1. Descriptive Statistics of the Big Five Traits

| Traits | Mean | SD | Alpha Coefficient |
|---|---|---|---|
| Extraversion | 3.27 | .86 | .89 |
| Agreeableness | 4.00 | .60 | .76 |
| Emotional Stability | 2.82 | .73 | .80 |
| Conscientiousness | 3.58 | .70 | .81 |
| Intellect/Imagination | 3.76 | .60 | .76 |

## 4     Analysis and Results

### 4.1    GAI Tool Use

Results suggest that totally 86% of students occasionally, frequently, or very frequently use GAI; About 70% of the students used GAI more than an hour per week; students used GAI for various purposes; About 35% of the students found GAI most useful for problem-solving; the most common concern is over-reliance on AI (44%), the next is the accuracy of the information generated (28%), and the third major concern is ethical concerns (17%); About 53% Students' preferred to use GAI tools over traditional resources for academic help, while 31% of students preferred the traditional resources. Approximately 49% of students learned about GAI from their friends or classmates, while only 12% learned about it from teachers or academic resources. While 53% of students received guidance from teachers/mentors on how to use GAI tools effectively, 39% did not report receiving guidance. Detailed statistics are presented in Appendices A[1] and B[1].

Principal component analysis with Promax rotation revealed three dimensions of the GAI uses: (a) GAI use for academic use, including frequency, duration, group use, peer influence on GAI use, and likelihood of future use; (b) Impact on learning and performance; (c) Ethical Concerns on using GAI. The three dimensions explained 75% of the original variance. Finally, the degree to which students prefer GAI over traditional resources for academic help is shown to be an independent construct. The five items measuring students' use of GAI for academic use have high internal consistency, Cronbach's alpha = .89. Therefore, we calculated the composite score, GAI Use, and used it in further analyses. However, the internal consistency of impact on learning and performance is rather low (alpha coefficient = .57). Therefore, we kept the two variables separate in the following analyses.

Based on the psychometrics validation, we focus on five GAI variables in the following analyses: *GAI Use, Impact on Performance, Impact on Learning, GAI Preference, and Ethical Concerns*. Table 2 reports the means of the five GAI variables. Appendix C includes more details, including the standard deviation of each variable.

Table 2. Means of the GAI Variables for Different Groups[1]

| Variable | Value | N | % | GAI Use | Impact on Performance | Impact on Learning | GAI Preference | Ethical Concern |
|---|---|---|---|---|---|---|---|---|
| Gender | Male | 181 | 49.9 | 3.08 | 3.94 | 3.14 | 3.28 | 2.7 |
|  | Female | 180 | 49.6 | 3.03 | 3.78 | 2.78 | 3.26 | 2.8 |
| Ethnicity | Asian/PI | 175 | 48.2 | 3.34 | 4.04 | 3.07 | 3.48 | 2.7 |
|  | Hispanic | 45 | 12.4 | 2.97 | 3.76 | 2.89 | 3.33 | 2.7 |
|  | Black | 23 | 6.3 | 2.71 | 4.17 | 3.09 | 3.17 | 2.6 |
|  | White | 95 | 26.2 | 2.69 | 3.58 | 2.94 | 2.99 | 2.9 |
|  | Multi | 22 | 6.1 | 2.88 | 3.76 | 2.41 | 2.82 | 2.9 |
| Major | Agriculture | 10 | 2.8 | 3.34 | 3.70 | 3.20 | 3.10 | 3.5 |
|  | Business | 194 | 53.4 | 3.07 | 3.89 | 2.99 | 2.76 | 3.3 |
|  | Engineering | 66 | 18.2 | 3.19 | 4.06 | 3.17 | 2.82 | 3.3 |
|  | Health | 26 | 7.2 | 2.57 | 3.81 | 2.77 | 2.85 | 2.6 |
|  | Science | 23 | 6.3 | 3.01 | 3.87 | 2.83 | 3.04 | 3.3 |
|  | Others | 34 | 9.4 | 3.85 | 3.76 | 2.59 | 2.94 | 3.1 |
| Grade | Fresh | 112 | 30.9 | 2.65 | 3.65 | 2.84 | 3.00 | 2.8 |
|  | Soph | 55 | 15.2 | 2.77 | 4.00 | 2.85 | 3.15 | 2.6 |
|  | Junior | 48 | 13.2 | 3.23 | 4.00 | 2.98 | 3.40 | 2.6 |
|  | Senior | 40 | 11.0 | 2.85 | 3.30 | 2.57 | 3.07 | 2.9 |
|  | Master | 85 | 23.4 | 3.62 | 4.20 | 3.16 | 3.74 | 2.9 |
|  | PhD | 21 | 5.8 | 3.51 | 4.43 | 3.81 | 3.19 | 3.1 |
| Native Speaker | Yes | 198 | 54.5 | 2.73 | 3.73 | 2.86 | 3.12 | 2.8 |
|  | No | 165 | 45.5 | 3.43 | 4.01 | 3.07 | 3.44 | 2.8 |
| Total |  | 363 |  | 3.05 | 3.86 | 2.96 | 3.26 | 2.8 |

*Note*: Small groups with sample sizes of fewer than 10 are not included in this table as the results are less stable.

Overall, students reported significantly higher GAI Use scores than their ethical concerns about it, $t(362) = 2.70$, $p < .007$, Cohen's $d = .14$; students felt that GAI has a more positive impact on their academic performance than their learning, $t(362) = 14.26$, $p < .001$, Cohen's $d = .75$.

### 4.2  Demographic Factors

To examine whether different student groups differ in their use and attitudes toward GAI, we conducted MANOVA. Results suggest that overall, (a) Male and female students did not differ in the five dimensions. (b) Students with different majors did not significantly differ either. (c) Students in different grades significantly differ in their responses (Fig. 1), Wilk's Lambda =.74, $F$ (25, 1305) = 4.46, $p < .001$. Univariate ANOVA reveals that overall students in higher grades tend to use GAI more than those in lower grades, perceive the positive impact of GAI on their learning and academic performance, and prefer to utilize GAI more frequently. However, students in different grades do not differ in their ethical concerns. (d) Students in different ethnic groups also differ in their responses on all the dimensions except for ethical concerns, Wilk's Lambda = .87, $F$ (20, 1148) = 2.55, $p < .001$. Univariate ANOVA shows that Asian students used GAI significantly more than White and Black students, felt that GAI had a more positive impact on their academic performance than White students, and preferred GAI more than traditional learning resources than White students. (e) Non-native English speakers differ significantly from native English speakers (Fig. 2), Wilk's Lambda = .86, $F$ (5, 357) = 11.74, $p < .001$. Univariate ANOVA shows that non-native English speakers significantly used GAI more than native English speakers ($F$ (1, 361) = 50.43, $p < .001$), perceived its positive impact on their academic performance more ($F$(1, 361) = 6.51, $p < .001$), and preferred GAI more than the traditional resources for help ($F$ (1, 361) = 5.55, $p = .019$).

We conducted multiple regressions with various GAI responses as dependent variables, grades, language status (dummy variable), and ethnicity (dummy variables with White as the reference group) as predictors so that we could further examine the association between students' GAI responses and their demographic information after controlling for other demographic information. The results showed similar patterns to what MANOVA showed: (a) GAI Use: The model significantly predicted GAI Use, $R^2$ = .23, $F$ (6, 356) = 17.94, p < .001. The higher the grade level ($\beta$ = .273, $t$ = 5.54, $p < .001$), non-native English speakers ($\beta$ = .249, $t$ = 4.77, $p < .001$), and Asian students ($\beta$ = .14, $t$ = 2.39, $p = .017$) use GAI more. (b) Impact on academic performance: The model significantly predicted a positive impact on academic performance: The model significantly predicted GAI impact on academic performance, $R^2$ = .07, $F$ (6, 356) = 4.23, p < .001. The higher the grade level ($\beta$ = .13, $t$ = 2.29, $p =.023$), Asian students ($\beta$ = .18, $t$ = 2.66, $p =.008$) and Black students ($\beta$ =.12, $t$ = 2.16, $p =.031$) reported more positive impact on their academic performance. (c) Preference of GAI over Traditional Resources: The model significantly predicted to what degree students prefer to use GAI over traditional resources for academic help, $R^2$ = .05, $F$ (6, 362) = 3.13, $p = .005$. The higher the grade level ($\beta$ = .11, $t$ = 2.07, $p = .04$) and Asian students ($\beta$ = .154, $t$ = 2.32, $p = .021$) prefer GAI more. (d) The model does not significantly predict the impact on learning and ethical concerns.

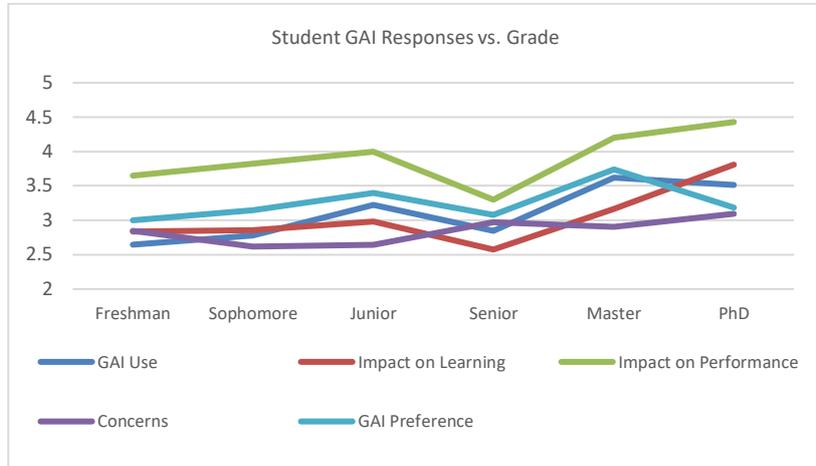

**Fig. 1.** Students' GAI responses vs. their grade levels

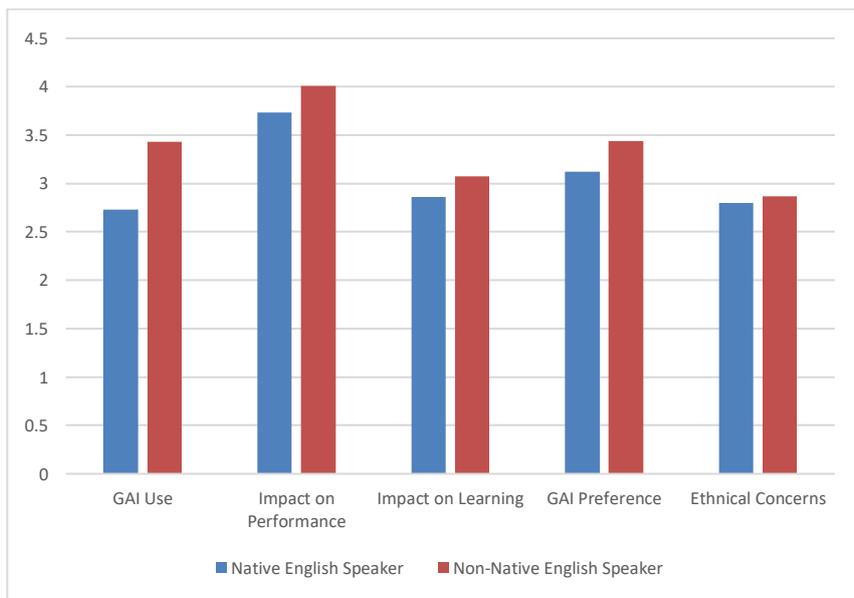

**Fig. 2.** Students' GAI responses vs. whether they are native English speakers.

### 4.3    Personality

Correlation coefficient analyses (Table 3) show that emotional stability positively correlated with a positive impact of GAI on performance, learning, and GAI preference, negatively correlated with ethical concerns. Extraversion is also positively correlated with GAI preference. In contrast, conscientiousness and intellect/imagination are negatively correlated with GAI use, and agreeableness is positively correlated with ethical concerns.

Table 3. Correlation Coefficients Between GAI Use and Personality

| | GAI Use | Impact on Performance | Impact on Learning | GAI Preference | Ethical Concern |
|---|---|---|---|---|---|
| Impact on Performance | .270** | | | | |
| Impact on Learning | .160** | .403** | | | |
| GAI Preference | .581** | .294** | .117* | | |
| Ethical Concern | -.153** | -.180** | -.166** | -.304** | |
| **Extraversion** | **0.076** | **0.085** | **0.047** | **.125*** | **-0.078** |
| **Agreeableness** | **-0.046** | **-0.08** | **-0.024** | **0.03** | **.112*** |
| **Emotional Stability** | **0.002** | **.108*** | **.183*** | **.104*** | **-.157*** |
| **Conscientiousness** | **-.125*** | **-0.013** | **0.065** | **-0.097** | **0.03** |
| **Intellect Imagination** | **-.111*** | **-0.055** | **0.099** | **-0.094** | **0.046** |

**. Correlation is significant at the 0.01 level (2-tailed).
*. Correlation is significant at the 0.05 level (2-tailed).
The correlation coefficients of the big five traits and GAI variables are bolded.

We further ran sequential regression to examine how GAI variables are associated with personality after students' demographic information is controlled for, specifically students' grade level and whether they are native English speakers. We did not include ethnicity in the sequential regression model because they are significantly correlated with language status and multiple personality traits. The regression results show (a) the Big Five traits significantly contributed to the prediction of GAI variables after controlling for the demographic information: the GAI Use ($\Delta R^2 = .032$, $F(5, 355) = 2.99$, $p = .012$), impact on performance ($\Delta R^2 = .032$, $F(5, 355) = 2.42$, $p = .035$), impact on learning ($\Delta R^2 = .045$, $F(5, 355) = 3.41$, $p = .005$), preference for GAI ($\Delta R^2 = .040$, $F(5, 355) = 3.04$, $p = .011$), and ethical concerns ($\Delta R^2 = .052$, $F(5, 355) = 3.93$, $p = .002$), b) GAI Use: individually, those who have higher conscientiousness reported lower GAI use, $\beta = -.122$, $t = -2.45$, $p = .015$. (b) Impact on academic performance: Those with higher extraversion reported a higher positive impact of GAI on their academic performance, $\beta = .11$, $t = 2.01$, $p = .046$, while those with higher agreeableness reported a lower positive impact of GAI on their academic performance, $\beta = -.12$, $t = -2.22$, $p = .027$. (c) Impact on learning experience: Those with higher emotional stability reported a higher positive impact on the learning experience, $\beta = .16$, $t = 3.16$, $p = .002$. (e) GAI Preference: Those with higher extraversion prefer GAI over the traditional resources more for academic help, $\beta = .12$, $t = 2.13$, $p = .034$; while those with high intellect/imagination prefer less GAI over the traditional resources, $\beta = -.11$, $t = -2.00$, $p = .047$. (d) Ethical concerns: Those with high agreeableness tend to have more ethical concerns about GAI, $\beta = .13$, $t = 2.43$, $p = .016$, while those with high extraversion ($\beta$

= -.11, $t$ = -1.99, $p$ = .047) and emotional stability ($\beta$ = -.17, $t$ = -3.17, $p$ = .002) have less ethical concerns about GAI use.

## 5. Discussion

The survey analysis results indicate that university students utilize GAI extensively for various purposes. However, they also express concerns, primarily regarding the accuracy of information and the potential for over-reliance on GAI. Despite these concerns, students' overall GAI use outweighs their concerns about its use. Additionally, students perceive GAI as having a more positive impact on their academic performance than on their learning, which aligns with their concerns about becoming overly dependent on the tool.

We compared the GAI use of different student subgroups. Students did not show significant differences in GAI use based on gender or academic major. However, students in higher grade levels were more likely to embrace GAI than those in lower grade levels. This trend may be attributed to multiple factors: (a) GAI tools, such as ChatGPT, are often discouraged in K-12 settings, whereas they are less restricted in higher education, particularly for graduate students. (b) Students in higher grade levels have more content knowledge and are therefore more proficient at utilizing GAI; for example, they can provide prompts that align with their desired outcomes. In contrast, younger students may lack the necessary understanding to formulate useful prompts that guide GAI in generating helpful information. For example, individuals who have sufficiently learned a programming language are more likely to utilize and benefit from GAI in coding than those with limited programming knowledge.

A comparison between native and non-native English speakers revealed that non-native English speakers used GAI significantly more than native English speakers. This trend may stem from the language barriers non-native speakers often encounter in academic settings where English is the dominant language, which likely drives them to rely more on GAI. For example, non-native English speakers may use GAI to proofread their essays or generate drafts more often than native English speakers. Interestingly, Asian students demonstrated the highest levels of GAI adoption, and both Asian and Black students reported a more positive impact of GAI on their academic performance compared to their White peers. Although a larger proportion of Asian students are non-native English speakers than white students, the disparity in GAI use between Asian and White students remains significant even after accounting for language proficiency.

The examination of personality measured by Big Five traits suggests that students' personality traits significantly predict their GAI use and perceptions, even after demographic information is controlled for. Students with higher extraversion reported a more positive impact on performance, more GAI preference, and less ethical concerns for GAI use academic assistance, which may be due to their tendency to seek external engagement and collaborative learning.

Students who are highly conscientious are less likely to use GAI. We suspect that their heightened sense of social responsibility and desire to avoid plagiarism or overreliance on GAI could make them more cautious about using GAI for coursework.

Students who have higher agreeableness did not perceive GAI as beneficial for their academic performance. Instead, they expressed greater concerns about the ethical implications of using GAI for academic work. Previous research has reported mixed findings on the relationship between agreeableness and GAI use, with both positive [5] and negative [6] effects observed. We suspect that the discrepancy in the agreeableness measure is due to the information to which the students are most exposed. If students receive more information that discourages them from using GAI and emphasizes the concerns associated with its use, students with high agreeableness may use GAI less and develop negative attitudes. On the other hand, if students are encouraged to use GAI, those with high agreeableness are more likely to adopt it. Given that different universities have varying policies on GAI use and instructors have differing attitudes toward GAI use, students with high agreeableness may be impacted differently in various contexts. Additionally, the outputs from GAI are prone to errors and often require multiple prompts and iterations. Highly agreeable students may be more likely to accept and use false information generated by GAI, only to later realize it is inaccurate, which can lead to mistrust in GAI.

Finally, students who have high extraversion and emotional stability show less ethical concerns about using GAI for academic work. Those with higher extraversion prefer GAI more than traditional resources, while students who have higher intellect/imagination prefer GAI less than traditional resources for academic help.

Overall, university students differ in their GAI use and attitudes. Both demographic information and personality account for these differences. GAI has great potential to improve not only academic performance but also learning. Students seemed to underestimate the benefit of GAI on learning compared to the reported benefit of GAI on academic performance.

It is noted that a large percentage of students reported not receiving guidance on GAI use from mentors or institutional figures. This gap highlights the need to guide students to effectively and appropriately use GAI so that GAI can essentially improve their learning.

The lack of guidance and regulation may also lead to fairness issues due to the differences in students' perceptions of GAI benefits and concerns, and consequently, their use of GAI. Institutional guidance should foster the ethical use of AI, ensuring that students can utilize GAI to effectively enhance their learning rather than avoiding it due to uncertainty or ethical concerns due to their demographic and personality differences.

## 6. Conclusion

Our study aligns with previous research, highlighting the various ways students integrate GAI into their academic practices, with note summarization, topic revision, and creative brainstorming being the most common uses. The findings suggest that GAI serves as a valuable tool for enhancing students' learning experiences, particularly in areas like problem-solving, coding, and technical writing. However, AI adoption is not uniform across all students; individual demographic information and personality traits influence how students use or perceive GAI tools.

Demographic factors, such as grade level and ethnicity, are also associated with the use of AI among university students. The higher the grade level, the more likely students are to use and prefer GAI. Non-native English speakers used GAI more than native English speakers, probably due to the strength of GAI in English writing. Asian and Black students were more likely to report positive impacts from GAI use on their academic performance.

The results also highlight that students' personalities play a crucial role in shaping their AI use patterns. Students who have high extraversion and emotional stability are more likely to adopt ChatGPT as a source of academic support. In contrast, students who have high conscientiousness use GAI less, possibly viewing AI as less compatible with their disciplined study habits or moral standards. Similarly, agreeable students exhibit heightened ethical apprehensions about AI's role in academia, possibly reflecting the overall discouragement from instructors of using GAI and their strong social desirability.

This study can be expanded in multiple directions: First, this study is limited to one-time cross-sectional survey data. Future research can collect longitudinal data on students' experiences and attitudes with GAI in their academic work. Investigating whether students' skepticism or acceptance of GAI evolves over time and responds to instructors' guidance, and how this process interacts with their personality traits would provide deeper insights into the dynamics of student engagement with GAI. Moreover, future research should interview students to discover the mechanism underneath the statistical relationship between GAI use and the influencing factors. Finally, empirical data are needed to study how to help students to use GAI to improve their learning, not just academic performance.